\documentclass[11pt]{article}
\usepackage{latexsym}  
\usepackage{amssymb}
\usepackage{graphicx}
\usepackage{amsmath}
\usepackage{mathrsfs}

\newcommand{\be}{\begin{equation}}
\newcommand{\ee}{\end{equation}}
\newcommand{\bea}{\begin{eqnarray}}
\newcommand{\eea}{\end{eqnarray}}
\newcommand{\bref}[1]{(\ref{#1})}
\newcommand{\bs}{\boldsymbol}

\newcommand{\bi}{\bibitem}
\newcommand{\pa}{\partial}

\newcommand{\bomega}{\boldsymbol{\Omega}}
\begin{document}
\begin{titlepage}
\vspace{4\baselineskip}
\begin{center}
{\Large\bf Addenda to General Spin Precession and Betatron Oscillation\\ in Storage Ring}
\end{center}
\vspace{1cm}
\begin{center}
{\large Takeshi Fukuyama
\footnote{E-mail:fukuyama@se.ritsumei.ac.jp}}
\end{center}
\vspace{0.2cm}
\begin{center}
{\small \it Research Center for Nuclear Physics (RCNP),
Osaka University, Ibaraki, Osaka, 567-0047, Japan}\\[.2cm]



\vskip 10mm
\end{center}
\vskip 10mm

\begin{abstract}
We give the geralized expression of spin precession of extended bunch particles having both anomalous magnetic and electric dipole moments in storage ring in higher order than the previous work and in the presence of ${\bf E}$ field as well as ${\bf B}$ field. These addenda are essential since some experiments consider the focusing field in the second order of the beam extent and in the presence of both ${\bf B}$ and ${\bf E}$ fields .
It is shown that some focusing fields with constant magnitude of the velocity considered in many literatures lead to the violation of self consistency.
\end{abstract}
\end{titlepage}
In the recent paper, we discussed general spin precession and betatron oscillation in storage ring \cite{fuku1}. "General" has the duplicate meanings. One is that the beam has a spread profile not only in vertical but also radial directions. Another is that it incorporates permanent electric dipole moment (EDM) as well as anomalous magnetic dipole moment (MDM). The measurement of both dipole moments are the smoking gun of the new physics beyond the standard model (SM) \cite{fuku3}. 
Recently the frequency shifts induced by magnetic frield gradients and ${\bf v}\times {\bf E}$ effects on the neutron EDM \cite{Pendlebury} and muon g-2 MDM \cite{Nouri} have been discussed.  
We emphasized the importance of 4-dimensional spin vector to discuss the spin of moving particles around the storage ring in the inertial frame.
Imaging the J-PARC muon g-2/EDM experiment \cite{J-PARC}, we considered ${\bf E}=0$ in the latter part of the previous paper.
In this paper, we consider the case ${\bf E}\neq 0$ but with the constant magnitude of the velocity like the case of BNL-E821 \cite{Bennett} and its successor at FNAL. Also taking the strength of focusing ${\bf B}$ and ${\bf E}$ fields into consideration, we make an approximation at higher order than the previous paper with respect to the small extents. 

The kinematical energy satisfies
\be
\frac{dE_{kin}}{dt}=e{\bf E}\cdot {\bf v}
\label{kinetic}
\ee
with
\be
E_{kin}\equiv \frac{m}{\sqrt{1-v^2}}\equiv \gamma m
\ee
and constant magnitude of the velocity requires
\be
{\bf E}\cdot{\bf v}=0.
\label{Ev}
\ee
Here and hereafter we use the $\hbar=c=1$ units.
In the presence of both ${\bf B}$ and ${\bf E}$, the Lorentz equation is given by
\be
\gamma m \frac{d{\bf v}}{dt}\equiv \gamma m\dot{{\bf v}}= e\left({\bf E}+{\bf v}\times {\bf B}-{\bf v}({\bf v}\cdot{\bf E})\right).
\label{Lorentz1}
\ee
The Lorentz equation is also expressed as
\be
\frac{d(\gamma m {\bf v})}{dt}=e\left({\bf E}+{\bf v}\times {\bf B}\right),
\ee
and under the constant magnitude of the velocity, the Lorentz equation becomes
\be
\gamma m \dot{{\bf v}}=e\left({\bf E}+{\bf v}\times {\bf B}\right).
\label{Lorentz2}
\ee
The generalized Thomas-Bargman-Michel-Telegdi (BMT) equation in the laboratory system is given by \cite{fuku2}
\bea
\bomega_s &=&-\frac{e}{m}\left[\left(a+\frac{1}{\gamma}\right){\bf B}-\frac{\gamma a}{\gamma+1}({\bf v}\cdot{\bf B}){\bf v}-\left(a+\frac{1}{\gamma+1}\right){\bf v}\times{\bf E}\right.\nonumber\\
&+&\left.\frac{\eta}{2}\left({\bf E}-\frac{\gamma}{\gamma+1}({\bf v}\cdot{\bf E}){\bf v}+{\bf v}\times {\bf B}\right)\right].
\label{Nelsonh}
\eea
One usually considers the spin motion relative to the beam direction. It goes from \bref{kinetic} and \bref{Lorentz1} that unit vector in direction of the velocity (momentum), ${\bf N}={\bf v}/v={\bf p}/p$ obeys
\be
\frac{d{\bf
N}}{dt}=\frac{\dot{\bf v}}{v}
-\frac{\bf v}{v^3}\left({\bf v}\cdot\dot{\bf v}\right)=\bomega_p\times{\bf N}, ~~~ \bomega_p=\frac{e}{m\gamma}\left(\frac{{\bf
N}\times{\bf E}}{v}-{\bf B}\right),
\ee 
Thus, the angular velocity of the spin rotation relative to the beam direction (prticle rest frame) is given by
\be
\bomega'=\bomega_s-\bomega_p=-\frac{e}{m}\left[a{\bf B}-\frac{\gamma a}{\gamma+1}({\bf v}\cdot{\bf B}){\bf v}-\left(a-\frac{1}{\gamma^2-1}\right){\bf v}\times{\bf E}+\frac{\eta}{2}\left({\bf E}-\frac{\gamma}{\gamma+1}({\bf v}\cdot{\bf E}){\bf v}+{\bf v}\times {\bf B}\right)\right].
\label{Nelsonf}
\ee
However, in the presence of the pitch correction, the particle rest frame is less convenient than the moving cylindrical frame, where particles move in horizontal plane on average and the deviation from it is approximated small (Fig.1). It is given by
\bea
\label{Nelsonf2}
&&\bomega=\bomega_s-\bomega_{pz}\nonumber\\
&&=-\frac{e}{m}\left[a{\bf B}-\frac{a\gamma}{\gamma+1}\boldsymbol{v}(\bs{v}\cdot{\bf B})+\left(\frac{1}{\gamma^2-1}-a\right)(\bs{v}\times\bs{E})\right.\\
&&\left. +\frac{1}{\gamma}\{\bs{B}_\parallel-\frac{1}{v^2}(\bs{v}\times\bs{E}_\parallel)\}+\frac{\eta}{2}(\bs{E}+\bs{v}\times\bs{B})\right].\nonumber
\eea
Here $\bomega_{pz}$ is the z-component of $\bomega_{p}$, and $\bs{B}_\parallel$ indicates the projected part of $\bs{B}$ on the averaged orbitting ($x-y$) plane, that is,
\be
\bs{B}=(B_x,B_y,B_z)=(\bs{B}_\parallel, B_z),~~\bs{E}=(\bs{E}_\parallel, E_z).
\ee 

\begin{figure}[h]
\begin{center}
\includegraphics[scale=2.0]{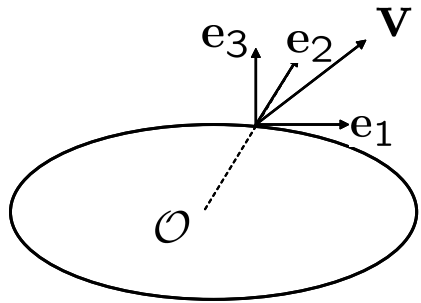}
\end{center}  
\caption{\label{fig:geom_config}
The configuration of betatron oscillation. $x({\bf e}_1),~y({\bf e}_2)$ coordinates are those projected on the averaged plane (horizontal place) of the storage ring. $z({\bf e}_3)$ is vertical to the horizontal plane. ${\bf v}$ is the particle velocity (emphasized to see the small pitch correction) which has both y, z components with main x component.}
\end{figure}

In this frame, 
\be
\frac{d{\bf e}_1}{dx}=-\frac{{\bf e}_2}{\rho},~~\frac{d{\bf e}_2}{dx}=\frac{{\bf e}_1}{\rho},~~\frac{d{\bf e}_3}{dx}=0.
\ee
Here ${\bf e}_1$ is the tangential unit vector (x-component) to the beam's averaged circular motion in the horizontal plane (spanned by x,y coordinates), 
${\bf e}_2$ is the radial unit vector (y-component) in the horizontal plane and ${\bf e}_3$ is the vertical unit vector (z-component) to the horizontal plane.
$\rho$ is the radius of the averaged circle and
\be
\frac{1}{\rho}=\left|\frac{eB_0}{\gamma mv}\right|.
\label{rho}
\ee 
Here we proceed to the bunch particles whose position ${\bf r}$ is given by
\be
{\bf r}=(\rho+y){\bf e}_2+z{\bf e}_3.
\label{profile}
\ee
Thus beam has (y,z) cross section relative to the averaged beam line.

So let us consider how Eq.\bref{Nelsonf2} is expressed in these profiles \cite{Yokoya}. The velocity of the profile of \bref{profile} is
\be
{\bf v}=\frac{d{\bf r}}{dt}=\frac{dx}{dt}\frac{d{\bf r}}{dx}=\frac{dx}{dt}\left[\left(1+\frac{y}{\rho}\right){\bf e}_1+y'{\bf e}_2+z'{\bf e}_3\right],
\label{velocity}
\ee
where $'$ indicates the derivative with respect to $x$. Therefore, the absolute value $v$ of ${\bf v}$ is given by
\be
v=\frac{dx}{dt}\sqrt{\left(1+\frac{y}{\rho}\right)^2+y'^2+z'^2}\equiv \frac{dx}{dt}{\mathscr N}.
\ee
In the previous paper we proceeded the arguments under the condition of ${\bf E}=0$. In this paper we relax this condition as ${\bf E}$ is weak but $\neq 0$ reserving the condition ${\bf E}\cdot{\bf v}=0$.

One of the typical examples of focusing fields of ${\bf B}$ and ${\bf E}$ are
\be
{\bf B}=\left(0, ~\frac{\pa B_z}{\pa y}z,~ B_0+\frac{\pa B_z}{\pa y}y\right), ~~~~{\bf E}=(0,~ Ky, ~Kz)
\label{focus}
\ee
They satisfy 
\be
\nabla\times {\bf B}=0, ~~~~\nabla\times {\bf E}=0. 
\label{rotation}
\ee
We assume $y/\rho,~z/\rho,~y',~z'$ are small quantities of same order $\epsilon$ and hereafter we take up to $O(\epsilon^2)$.
Hereafter we give the approximation formulae for the general ${\bf B}$ and ${\bf E}$, not assuming the special form of \bref{focus} until we will argue at \bref{constraint1} and thereafter.

Here we list the useful relations and approximations
\bea
&&\frac{1}{\gamma v^2}=\frac{\gamma}{\gamma^2-1}, \nonumber\\
&&\frac{1}{{\mathscr N}}=1-\frac{y}{\rho}+\frac{y^2}{\rho^2}-\frac{{y'}^2+{z'}^2}{2}+O(\epsilon^3), \\
&&\frac{1}{{\mathscr N}^2}=1-\frac{2y}{\rho}+\frac{3y^2}{\rho^2}-{y'}^2-{z'}^2+O(\epsilon^3), \nonumber
\eea
and the divergence and rotation of a general vector ${\bf B}$ (and ${\bf E}$)
in this approximation are
\bea
\label{veq1}
&&{\bf v}\cdot {\bf B}=\frac{v}{{\mathscr N}}\left((1+\frac{y}{\rho})B_x+y'B_y+z'B_z\right)\nonumber\\
&&\approx v\left((1-\frac{y'^2+z'^2}{2})B_x+(1-\frac{y}{\rho})(y'B_y+z'B_z)\right)+O(\epsilon^3),\\
\label{veq2}
&& ({\bf v}\cdot {\bf B}){\bf v}\approx v^2\left[\left((1-(y'^2+z'^2))B_x+(1-\frac{y}{\rho})(y'B_y+z'B_z)\right){\bf e}_1+y'\left((1-\frac{y}{\rho})B_x+y'B_y+z'B_z\right){\bf e}_2\right. \nonumber\\
&&\left.+z'\left((1-\frac{y}{\rho})B_x+y'B_y+z'B_z\right){\bf e}_3\right],\\
&&{\bf v}\times {\bf B}=\frac{v}{{\mathscr N}}\left[\left(y'B_z-z'B_y\right){\bf e}_1+\left(z'B_x-(1+\frac{y}{\rho})B_z\right){\bf e}_2+\left((1+\frac{y}{\rho})B_y-y'B_x\right){\bf e}_3\right]\nonumber\\
\label{veq3}
&&\approx v\left[(1-\frac{y}{\rho})(y'B_z-z'B_y){\bf e}_1+\left((1-\frac{y}{\rho})z'B_x-(1-\frac{y'^2+z'^2}{2})B_z\right){\bf e}_2\right.\\
&&\left.+\left((1-\frac{y'^2+z'^2}{2})B_y-y'(1-\frac{y}{\rho})B_x\right){\bf e}_3\right]. \nonumber
\eea

Substituting \bref{veq1}-\bref{veq3} into \bref{Nelsonf2}, we obtain
\bea
\label{Nelsonf4}
&&\bomega=-\frac{e}{m}\left[\left\{\frac{1}{\gamma}(a+1)B_x{\bf e}_1+\left((a+\frac{1}{\gamma})B_y-\frac{\eta}{2}vB_z\right){\bf e}_2+\left(aB_z+\frac{\eta}{2}vB_y\right){\bf e}_3\right\}\right. \nonumber\\ 
&&+\left\{\left(a(1-\frac{1}{\gamma})((y'^2+z'^2)B_x-(1-\frac{y}{\rho})(y'B_y+z'B_z))-(a+\frac{1}{\gamma+1})v(1-\frac{y}{\rho})(y'E_z-z'E_y)\right. \right.\nonumber\\
&&\left.\left.+\frac{\eta}{2}(E_x+v(1-\frac{y}{\rho})(y'B_z-z'B_y))\right){\bf e}_1\right.\\
&&\left.\left.+\left(-a(1-\frac{1}{\gamma})((1-\frac{y}{\rho})B_x+y'B_y+z'B_z)y-(a+\frac{1}{\gamma+1})v((1-\frac{y}{\rho})z'E_x-(1-\frac{y'^2+z'^2}{2})E_z)\right.\right.\right.\nonumber\\
&&\left.\left.\left.+\frac{\eta}{2}(E_y+v((1-\frac{y}{\rho})z'B_x+\frac{y'^2+z'^2}{2}B_z))\right){\bf e}_2\right.\right.\nonumber\\
&&\left.\left.+\left(-a(1-\frac{1}{\gamma})((1-\frac{y}{\rho})B_x+y'B_y+z'B_z)z'+(\frac{1}{\gamma^2-1}-a)v((1-\frac{y'^2+z'^2}{2})E_y-y'(1-\frac{y}{\rho})E_x)\right.\right.\right.\nonumber\\
&&\left.\left.\left.+\frac{\eta}{2}(E_z-v(\frac{y'^2+z'^2}{2}B_y+y'(1-\frac{y}{\rho})B_x))\right){\bf e}_3\right\}\right].\nonumber
\eea
Here the division of the first $\{...\}$ term and the second $\{...\}$ term is simply convenient for the reference to the previous paper \cite{fuku1}.

Next, we must study the equation of motions of $y$ and $z$ to see the time variation of \bref{Nelsonf4}.
Constant magnitude of velocity indicates that
\be
\frac{dv}{dt}=\frac{d^2x}{dt^2}{\mathscr N}+{\mathscr N}'\left(\frac{dx}{dt}\right)^2=0.
\ee
and, therefore,
\bea
\label{acceleration1}
&&\frac{d{\bf v}}{dt}=v^2\left[\left\{(1-\frac{2y}{\rho})\frac{y'}{\rho}-y'y''-z'z''\right\}{\bf e}_1+\left\{(1-\frac{2y}{\rho})y''-\frac{1}{\rho}(1-\frac{y}{\rho}+\frac{y^2}{\rho^2}-z'^2)\right\}{\bf e}_2 \right.\nonumber\\
&&\left. -\left\{\frac{y'z'}{\rho}-z''(1-\frac{2y}{\rho})\right\}{\bf e}_3\right].
\eea
Hereafter we consider the case of \bref{focus}.
This form together with \bref{Ev} and \bref{velocity} gives
\be
y^2+z^2=const.
\label{constraint1}
\ee
Consequently, in the $O(\epsilon^2)$ approximation, it goes from \bref{focus}, \bref{acceleration1} and from the Lorentz equation \bref{Lorentz2} that
\bea\label{eq1}
&& y'\left(y''+\frac{1-n}{\rho^2}y\right)+z'\left(z''+\frac{n}{\rho^2}z\right)=0,\\
&&y''+\frac{1-n_y}{\rho^2}y=\frac{1}{\rho}\left(2yy'+\frac{y^2}{\rho^2}+\frac{y'^2-z'^2}{2}\right),\\
&&z''+\frac{n_z}{\rho^2}z=\frac{1}{\rho}(2yz''+y'z')
\label{eq3}.
\eea
Here
\be
n\equiv -\frac{\rho}{B_0}\frac{\partial B_z}{\partial y},~~n_y\equiv n\left(1-\frac{K}{v\partial B_z/\partial y}\right),~~n_z\equiv n\left(1+\frac{K}{v\partial B_z/\partial y}\right).
\ee
Thus the well knowm Hill's equations,
\bea
&&y''+\frac{1-n}{\rho^2}y=0, \label{eqmotion1}\\
&&z''+\frac{n}{\rho^2}z=0,
\label{eqmotion2}
\eea
are consistent in $O(\epsilon)$ and in the absence of ${\bf E}$.
Obviously, \bref{constraint1} and \bref{eq1}-\bref{eq3} are incompatible in $O(\epsilon^2)$. However, we should remind that the form of ${\bf E}$ may be in general
\be
{\bf E}=(0, ~f(y), ~g(z))~~\footnote{Here, electric field at the average orbit is assumed to be zero.}
\ee
with arbirary functions of $f$ and $g$, which is restricted only by \bref{Ev} and \bref{rotation}. For instance, if we set
\be
{\bf E}=(0,~Ky,~-Kz),
\ee
in place of \bref{focus}, then
\be
n_z \to n_y.
\ee
However, \bref{constraint1} changes to
\be 
y^2-z^2=const.
\label{constraint2}
\ee

In conclusion, there is no problem on \bref{Nelsonf4}. The problem lies in the consistency of the equations of motion of $y$ and $z$ with the constant magnitude of velocity in the presence of ${\bf E}$. We need some careful agjustment mechanism for the consistency, which has not been considered seriously so far.
\section*{Acknowledgements}
We are greatly indebted to Dr. N. Saito, Dr. T. Mibe and Dr. T. Shimomura for useful discussions and hospitality at J-PARC. 
The work of T.F. is supported in part by Grant-in-Aid for Science Research from the Ministry of Education, Science and Culture (No.~26247036).

\end{document}